\documentclass[12pt]{article}

\usepackage{newtxtext,newtxmath}

\usepackage{graphicx}

\usepackage[letterpaper,margin=1in]{geometry}

\renewenvironment{abstract}
	{\quotation}
	{\endquotation}

\date{}

\makeatletter
\renewcommand{\fnum@figure}{\textbf{Figure \thefigure}}
\renewcommand{\fnum@table}{\textbf{Table \thetable}}
\makeatother

\usepackage{scicite}

\usepackage{url}

\def\scititle{
	Manipulating the direction of turbulent energy flux via tensor geometry in a two-dimensional flow
}
\title{\bfseries \boldmath \scititle}

\author{
	Xinyu~Si$^{1}$,
	Filippo~De~Lillo$^{3}$,
    Guido~Boffetta$^{3}$,
	Lei~Fang$^{1,2\ast}$\and
	\small$^{1}$Department of Civil and Environmental Engineering, University of Pittsburgh, \and
    \small Pittsburgh, 15261, Pennsylvania, USA.\and
    \small$^{2}$Department of Mechanical Engineering and Materials Science, University of Pittsburgh, \and 
    \small Pittsburgh, 15261, Pennsylvania, USA.\and
	\small$^{3}$Dipartimento di Fisica Generale and Istituto Nazionale di Fisica della Materia, Universitá di Torino, \and 
    \small Torino, 10125, Italy.\\ [1ex]
	\small$^\ast$Corresponding author. Email: lei.fang@pitt.edu
}

\begin{document} 

\maketitle

\begin{abstract} \bfseries \boldmath
In turbulent flows, energy flux refers to the transfer of kinetic energy across different scales of motion, a concept that is a cornerstone of turbulence theory. The direction of net energy flux is prescribed by the dimensionality of the fluid system. According to Kolmogorov's 1941 scaling theory \cite{kolmogorov1941local}, three-dimensional turbulence has a net energy flux toward smaller length scales, while in two-dimensional turbulence, energy transfers toward larger scales, as described in Kraichnan and Batchelor's seminal works \cite{kraichnan1967inertial,batchelor1969computation}. Manipulating energy flux across different scales with localized physical perturbations in flow systems is a formidable task because the energy at any scale is not localized in physical space. Here, we report a theoretical framework that enables the manipulation of energy flux direction in weakly turbulent flows. Based on this framework, we successfully manipulated a flow system to achieve the desired directions of net energy flux through both electromagnetically driven thin-layer flow experiments and direct numerical simulations. Significantly, we generated a type of turbulent flow that has never been produced before—two-dimensional Navier-Stokes turbulence with a net forward energy flux. Apart from theoretical interest, we discuss how our theoretical framework can have profound applications and implications in natural and engineered systems across length scale ranges from $10^{-3}$ to $10^{6}$ m, including enhanced mixing of microfluidic devices, biologically generated turbulence for a better understanding of the biogeochemical structure of water columns, breaking persistent coastal transport barriers for better coastal ecological health, and ocean energy budget in facing of climate change.

\end{abstract}

\noindent

Turbulence governs the motion of many fluid systems, including the oceans and atmosphere, and serves as an efficient mechanism for mixing substances. From a theoretical point of view, turbulence is the quintessential example of a non-linear system far from equilibrium with many degrees of freedom. Therefore, any advancement in understanding turbulence has significant implications and applications across multiple scientific fields.

Navier-Stokes (NS) turbulence is characterized by energy flux between different scales of motion. The direction of the net energy flux is predetermined by the dimensionality of the flow \cite{kolmogorov1941local,kolmogorov1962refinement,kraichnan1967inertial,pope_2000}. 
Heuristically, in three-dimensional (3D) turbulence, energy injected at macroscopic scales generates large eddies that break down into progressively smaller ones. This energy transfer towards smaller scales, known as forward energy flux, is eventually halted by viscous dissipation \cite{richardson1920supply,kolmogorov1941local}.
In contrast, in two-dimensional (2D) turbulence, energy is transferred from the scales where it is injected to larger scales—a process known as inverse energy flux. This energy is then either dissipated or accumulated at the largest available scale \cite{xia2011upscale,fang2017multiple,fang2021spectral} (Fig. \ref{fig1}a).


Here, we study an intriguing yet pragmatic question of whether the direction of net turbulent energy flux can be manipulated by a suitable forcing scheme.
Our manipulation approach is based on a simple observation that the turbulent cascade process can be recast into a mechanical process \cite{fang2016advection} where stress (analogous to force) and rate of strain (analogous to displacement) at different scales of motion can work with or against each other to generate positive or negative work between scales. In 2D turbulence, both stress and rate of strain are represented as second-order tensors. When the stress tensor aligns with the rate of strain tensor, small scales do work on larger scales, resulting in an inverse energy flux. Conversely, forward energy flux emerges when these two tensors are perpendicular (Fig. \ref{fig1}). This mechanical picture immediately underscores the critical role of geometry in determining the direction of spectral energy flux. The key to manipulating energy flux lies in controlling the alignment between these two tensors. If this intuitive framework holds, it could enable the generation of novel types of NS turbulence—specifically, 3D turbulence with a net inverse energy flux and 2D turbulence with a net forward energy flux. In this paper, we focus on manipulating 2D flow and show the successful control of net energy flux direction through both electromagnetically driven thin-layer flow experiments and direct numerical simulations. The framework can be extended to 3D flows as well.



\section{Theoretical framework of tensor alignment}
Filtering is an archetypal method for examining interactions between different scales in a nonlinear system.
By applying a filter to a nonlinear equation at a given length scale, the nonlinearity produces new terms in the filtered equation that capture the interaction between the degrees of freedom that are retained and those that are removed. In other words, these new terms act as source or sink terms for the remaining degrees of freedom. For example, applying a low-pass filter, i.e., removing scales of motion that are smaller than a certain cutoff length scale (L), to the NS equations introduces the subgrid-scale stress $\tau_{ij}^{(\rm L)} = (u_i u_j)^{(\rm L)} - u_i^{(\rm L)}u_j^{(\rm L)}$ into the filtered NS equations, where $u_i$ is the $i^{th}$ component of the fluctuating velocity. This stress term depicts the momentum transfer across the length scale L. 
Similarly, inspecting the equation of motion for filtered kinetic energy (${\rm E}^{(\rm L)} = \frac{1}{2}u_i^{(\rm L)}u_i^{(\rm L)}$) yields a spectral energy flux term $\Pi^{(\rm L)} = - \tau_{ij}^{(\rm L)} s_{ij}^{(\rm L)}$, representing the energy flux between unresolved and resolved scales, where $s_{ij}^{(\rm L)} = (1/2)(\partial u_i^{(\rm L)}/\partial x_j + \partial u_j^{(\rm L)}/\partial x_i)$ is the filtered rate of strain (see Methods).
Recalling the analogy introduced earlier, the $\tau_{ij}^{(\rm L)}$ is analogous to force and $s_{ij}^{(\rm L)}$ is analogous to displacement. The inner product between these terms determines the work done from filtered (smaller) scales to retained (larger) scales through length scale L, which represents the spectral energy flux between scales of motion. Manipulating $\Pi^{(\rm L)}$ is the primary goal of this study. 

This interpretation highlights the critical importance of geometric alignment between the two tensors $\tau_{ij}^{(\rm L)}$ and $s_{ij}^{(\rm L)}$ (Fig. \ref{fig1}). When $\tau_{ij}^{(\rm L)}$ and $s_{ij}^{(\rm L)}$ are aligned, $\Pi^{(\rm L)} < 0$, indicating inverse energy flux toward larger length scales. Conversely, when $\tau_{ij}^{(\rm L)}$ and $s_{ij}^{(\rm L)}$ are perpendicular, $\Pi^{(\rm L)} > 0$, indicating forward energy flux toward smaller length scales (Fig. \ref{fig1}).
Furthermore, $\Pi^{(\rm L)}$ can be reexpressed as a function that depends on the geometric alignment between the eigenframes of $\tau_{ij}^{(\rm L)}$ and $s_{ij}^{(\rm L)}$. In 2D flow, this relationship is described by the following equation \cite{liao2014geometry,fang2016advection}: 
\begin{equation} \label{eqn: tensor geometry}
    \Pi^{(\rm L)} = -2 \gamma \sigma \textrm{cos}(2\theta^{(\rm L)}),
\end{equation}
where $\sigma$ and $\gamma$ are the largest eigenvalues of the rate of strain and the deviatoric part of stress tensors, and $\theta^{(\rm L)}$ is the angle between the corresponding (extensional) eigenvectors $\hat{\sigma}$ and $\hat{\gamma}$.
It is then clear that the alignment of the stress and the rate of strain tensor can determine not only the magnitude but also the direction of the energy flux. When $\theta^{(\rm L)} < \pi /4$, energy fluxes to larger scales generating inverse energy flux; when $\theta^{(\rm L)} > \pi /4$, energy fluxes to smaller scales, resulting in forward energy flux. No net energy flux occurs when $\theta^{(\rm L)} = \pi /4$. In typical isotropic 2D turbulent flows, the alignment between stress and rate of strain tensor is self-organized, leading to a net inverse energy flux. Previous researchers have proposed treating $\eta = \textrm{cos}(2\theta^{(\rm L)})$ as a measure of the efficiency of energy flux between scales \cite{fang2016advection}. It was found that the efficiency in typical isotropic 2D turbulent flow is relatively low, with an $\eta$ of only 27\%, as reported in previous experiments \cite{fang2016advection}. 

In principle, by generating a background flow with an ordered rate of strain and perturbing it with directionally biased stresses, we can control the tensor geometry between the stress and rate of strain tensors, thereby manipulating the efficiency of the net energy flux based on Eqn. \ref{eqn: tensor geometry}.
In this study, we selected hydrodynamic shear as the background flow, establishing a well-organized large-scale rate of strain orientation (Fig. \ref{fig2}b), and perturbed it with a directionally biased monopole-like perturbation (Fig. \ref{fig2}c). The direction of $\hat{\gamma}$ from the monopole-like perturbation is found to align with the direction of the applied monopole forces (see Methods). Consequently, by controlling the mechanical angle $\theta$ between the direction of $\hat{\sigma}$, associated with the background shear flow, and the direction of the monopole forces, we can significantly manipulate the direction of the spectral energy flux.



\section{Experiments and numerical simulation of energy flux manipulation}
To apply this theoretical framework for manipulating energy flux, we conducted experiments using an electromagnetically driven thin-layer flow system (Fig. \ref{fig2}a) \cite{fang2017multiple,si2024interaction,si2024biologically}. We generated a steady shear flow to establish a well-ordered large-scale rate of strain via the Lorentz body force that arose from the interaction between the magnetic field produced by two stripes of magnets with opposite polarities and a direct current passing through the electrolyte layer (Fig. \ref{fig2}b). 
The physical perturbation was introduced using a 5 by 5 grid of rods driven by a programmable linear actuator at a velocity of 1 cm/s in a forward-and-back manner (Fig. \ref{fig2}c). The flow was then recorded and analyzed using a particle tracking velocimetry (PTV) algorithm \cite{ouellette2006quantitative} (see Methods).

The numerical simulations were performed using a standard fully dealiased pseudospectral code
\cite{boyd2001chebyshev,valadao2024spectrum}. NS equations were integrated on a 2D domain with a second-order Runge-Kutta temporal scheme. The linear hydrodynamic shear was generated by simulating a Couette flow between two walls. Local physical perturbation was applied via a 5 by 5 array of force monopoles. The strength of monopoles had a pulsating force varying sinusoidally (see Methods).


\section{The manipulated net spectral energy flux}
Our experimental and simulation results of energy flux manipulation are summarized in Fig. \ref{fig3}, where a significant correlation between the controlled mechanical angle ($\theta$) and the measured tensor alignment angle ($\theta^{(\rm L)}$) was observed.
In our experiments, we conducted three control cases. The first involved pure shear flow without perturbation. In the second case, a static rod array was introduced into the shear flow, ensuring that any changes in tensor geometry were not due to the rod array acting as a new boundary condition. The third control case was the rod array moving in a quiescent fluid to rule out the possibility that the energy flux manipulation was due to the rod array moving alone. 
As shown in Fig. \ref{fig3}a, in all three control cases, $\theta^{(\rm L)}$ was symmetrically distributed around $\pi /4$, resulting in an efficiency $\eta$ close to zero. Consequently, we observed only a relatively weak spectral energy flux in these control cases (Fig. \ref{fig3}c). 

When we aligned the added stress with the background rate of strain ($\theta \approx 0$; $\eta$ = -1), we observed a salient shift of $\theta^{(\rm L)}$ toward 0 (Fig. \ref{fig1}a). Similarly, we observed a significant shift toward $\theta^{(\rm L)}$ = $\pi/2$ as we applied the added stress perpendicularly with the background rate of strain ($\theta \approx \pi/2$; $\eta$ = 1). As our manipulation set $\theta$ to approximately $\pi/4$, $\theta^{(\rm L)}$ was symmetrically distributed around $\pi /4$, resulting in only a small net energy flux between scales (Fig. \ref{fig3}a and h). 
The direct numerical simulations allowed for fine-tuning the direction of the monopole array. In the inset of Fig. \ref{fig3}j, we present the energy flux as a function of different mechanical angles $\theta$. We see that the energy flux varied with $\theta$ in a sinusoidal manner that reflected the form of Eqn. \ref{eqn: tensor geometry}. Theoretically, the maximum inverse energy flux should occur when $\theta^{(\rm L)}$ = 0, and the maximum forward energy flux will emerge when $\theta^{(\rm L)}$ = $\pi/2$. In our observations, the maximum inverse and maximum forward angle alignments occurred at $\theta = \pi/16$ and $\theta = \pi/2$, respectively. The slight discrepancy between the optimal $\theta$ and optimal $\theta^{(\rm L)}$ for maximum inverse energy flux is likely due to the engineered tensor alignment being slightly altered during the coupling, an inherent non-linear process, between the physical perturbation and the background flow. 

We calculated the energy flux between scales based on the measured stress and rate of strain tenors. In Fig. \ref{fig3}b and i, we present the time series of the spatially averaged energy flux. Although the time series of energy flux correlated with the forward-and-back motion of the rod array in experiments and with the blinking of the monopoles in simulations, the energy flux directions remained consistent with the manipulated geometric alignments. We also calculated the net energy flux across different cutoff scales (Fig. \ref{fig3}c and j). Consistent with the tensor geometry statistics, the motion of the rod and monopole arrays significantly influenced the direction of the energy flux by introducing directionally biased small-scale stresses.


A further observable related to the direction of energy transfer is 
 the third-order longitudinal structure function $S_3(r) = \langle
[\Delta_r \mathbf{u}\cdot\mathbf{\hat{e}_l}]^3 \rangle$, where
$\mathbf{\hat{e}_l}$ is the unit vector in the longitudinal direction and
$\Delta_r \mathbf{u} = \mathbf{u}(\mathbf{x} + {\rm r}\mathbf{\hat{e}_l}) -
\mathbf{u}(\mathbf{x})$ is the velocity difference over displacement r. While the filtering approach in Eqn. \ref{eqn: tensor geometry} accesses different scales by spatial filtering, $S_3(r)$ encodes the information on the dynamics at each scale via the statistics of the velocity difference at the corresponding displacements in physical space. As shown in Fig. \ref{fig3}d and k, the third-order structure function changes sign with $\theta$. This can be interpreted in view of well-known results valid for the inertial range of large-Reynolds-number turbulent flows. In that case, one can show that $S_3 = -C \epsilon {\rm r}$, where $\epsilon$ is the (positive) energy
dissipation rate while $C$ is a constant whose sign depends on the direction of the cascade. In 3D flows (where the  flux is positive) $C =\frac{4}{5}$ \cite{kolmogorov1991dissipation}, while $C = -\frac{3}{2}$ \cite{alexakis2018cascades} in 2D flow, where the energy flux is negative and an inverse, upscale energy 
cascade is observed. 
Although at our relatively low Reynolds numbers, the scaling results do not apply, one can expect $S_3<0$ for a direct energy flux and $S_3>0$ for an inverse one, which is consistent with the observation. Therefore, the sign of $S_3$ provides an additional signature of the direction of
spectral energy flux, complementing the filtering results.

Significantly, we have both experimentally and numerically produced 2D weak turbulence with net forward energy flux-a type of NS turbulence that has never been generated before (Fig. \ref{fig3}e and l). This is particularly noteworthy because traditional 2D turbulence, as predicted by Kraichnan \cite{kraichnan1967inertial}, exhibits a net inverse energy flux. The creation of this new type of turbulence provides a unique opportunity to compare it with its traditional counterpart, potentially deepening our understanding of the turbulent cascade process. From an application perspective, reversing the natural direction of energy flux may induce profound kinematic and dynamical differences that may not only enhance our understanding of natural processes but also improve our ability to control engineered systems. 



\section{Applications and implication in natural and engineered systems}
Our analysis demonstrates that directionally biased physical perturbation can couple with the background flow, causing distinct yet predictable directions of spectral energy flux. Directionally biased physical perturbations are prevalent in both natural and engineered systems. Therefore, our results have broad applications and implications in both natural and engineered systems, spanning length scales from millimeters in microfluidic mixers to hundreds of kilometers in geophysical flows.

On the millimeter scale, microfluidic mixers often suffer from poor mixing \cite{stroock2002chaotic,stone2004engineering}. Our findings offer valuable insights into addressing this issue. By engineering the flow in microfluidic mixers at a low Reynolds number to induce forward energy flux, it is possible to generate smaller scales of motion, thereby enhancing mixing efficiency.

Biologically generated ocean mixing plays a crucial role in understanding the biogeochemical structure of the water column in climatically important regions of the ocean \cite{houghton2018vertically,kunze2006observations,katija2012biogenic}. Contrary to traditional belief, a recent study has shown that a swimmer's ability to mix the local flow is not an immutable trait but varies depending on the swimmer's alignment relative to local shear. The study demonstrated that flows generated by a group of swimmers can couple with background flows to enhance mixing \cite{si2024interaction}. Moreover, the interaction between the directionally biased stress from a swimmer and a background hydrodynamic shear can induce significant differences in spectral energy transfer properties and modify the strength of background hydrodynamic shear \cite{si2024biologically}. Therefore, the coupling between directionally biased stress from swimmers and background flow is of great importance in understanding the impact of biologically generated turbulence on ocean mixing.

In coastal oceans, Lagrangian Coherent Structures (LCSs), which can span several kilometers, act as transport barriers in geophysical flows, hindering effective mixing in coastal areas and potentially contributing to the formation of ocean forbidden zones \cite{olascoaga2006persistent,fang2018influence}. Disrupting these LCSs in coastal regions could alleviate these forbidden zones and improve the health of coastal ecosystems. Our theoretical framework offers a method to engineer optimal small-scale stress that couples with the background flow to enhance forward energy flux. The enhanced forward energy flux will dump energy that sustains the large-scale LCSs to smaller scales, where eventually it can be dissipated by viscosity. In the Methods section, we present a theoretical estimation demonstrating the feasibility of manipulating LCSs. This estimation suggests that it is possible to substantially influence LCSs using only 0.05\% of the energy that sustains them.

In geophysical systems, wind stresses consistently do positive or negative work to facilitate energy exchange between atmospheric and oceanic systems \cite{dewar1987some,renault2016control,rai2021scale}. Beyond this traditional first-order view of energy exchange, our results indicate that a profound second-order effect may arise when local wind stresses act as biased stresses. These biased wind stresses could interact with the rate of strain of the oceanic flow, leading to distinct directions of energy flux among different scales of motion in oceanic flows. In the context of climate change, alterations in the flow patterns of either atmospheric or oceanic systems can influence not only the energy exchange between these systems but also the direction of energy flux within the oceanic flow system. Our findings provide valuable insights into the profound second-order effects of geophysical flow in the face of climate change.

\section{Conclusion}
We have developed a theoretical framework for manipulating the direction of spectral energy flux through tensor geometry. This theoretical framework was demonstrated through the successful manipulation of spectral energy flux of 2D flow in both experiments and simulations. Beyond its theoretical significance, our framework has profound applications and implications for natural and engineered systems ranging from microfluidic mixers and biologically generated turbulence to geophysical flows. 


\clearpage

\begin{figure}[p]
    \centering
    \includegraphics[width=\textwidth]{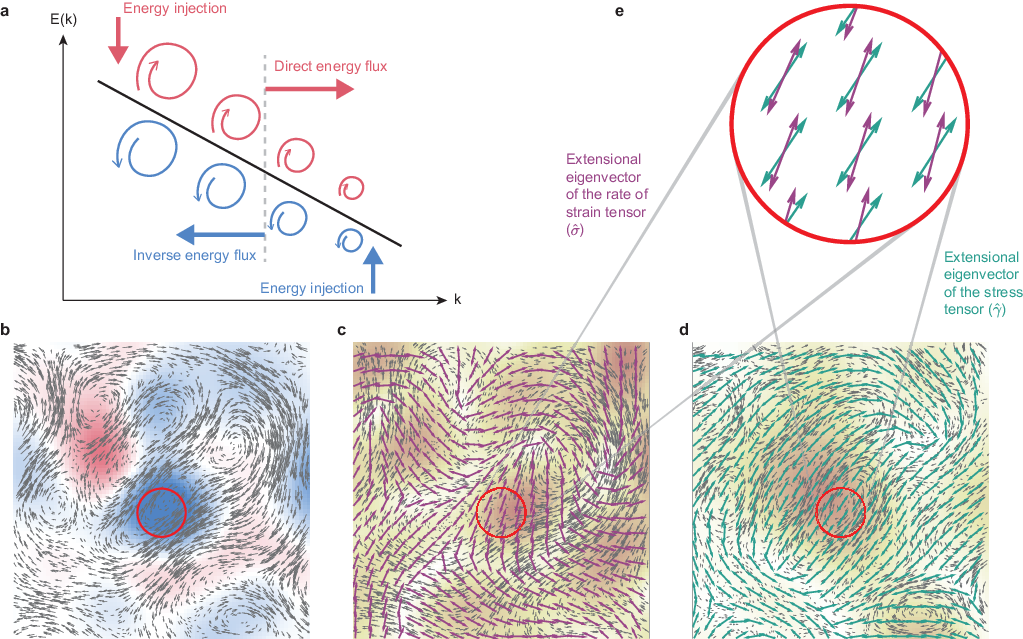} 
\end{figure}
\begin{figure}
  \caption{\textbf{Spectral energy flux and tensor geometry.} \textbf{a}, A schematic representation of spectral energy flux in turbulence. For forward energy flux (red), energy injected at large scales cascades to progressively smaller scales until it is dissipated by viscous forces. For inverse energy flux (blue), energy is injected at small scales and then transferred to progressively larger scales. This energy is either dissipated or piles up at the largest scale available within the system, defined by the system's size. These processes are quantitatively described by the energy spectrum E(k), which denotes the distribution of kinetic energy across modes with wavenumber k = 2$\pi$/L. 
    \textbf{b}, Instantaneous velocity field (gray arrows) overlaid on a spectral energy flux map for 2D weakly turbulent flow. Consistent with the color scheme in \textbf{a}, red color represents forward energy flux, and blue stands for inverse energy flux. The intensity of the color indicates the magnitude. \textbf{c}, Large-scale velocity $u_i^{(\rm L)}$ (gray arrows) for the same 2D turbulent flow, with L/W = 0.8, where W is half of the domain size. Purple double-headed arrows indicate the local direction of $\hat{\sigma}$, and the background color shows the magnitude of $\sigma$. \textbf{d}, Small-scale velocity $u_i - u_i^{(\rm L)}$ (gray arrows) for the same 2D turbulent flow. Green double-headed arrows indicate the local direction of $\hat{\gamma}$, and the background color shows the magnitude of $\gamma$. \textbf{e}, A zoomed-in view of tensor geometry, showing the alignment between the extensional eigenvectors of the rate of strain tensor (purple) and of the stress tensor (green). The local spectral energy flux depends on the tensor geometry, as described by Eqn.~\ref{eqn: tensor geometry}. The alignment between these two eigenvectors in the circled region is consistent with the energy flux direction in \textbf{b}.}
    \label{fig1}
\end{figure}

\begin{figure}[p]
    \centering
    \includegraphics[width=\textwidth]{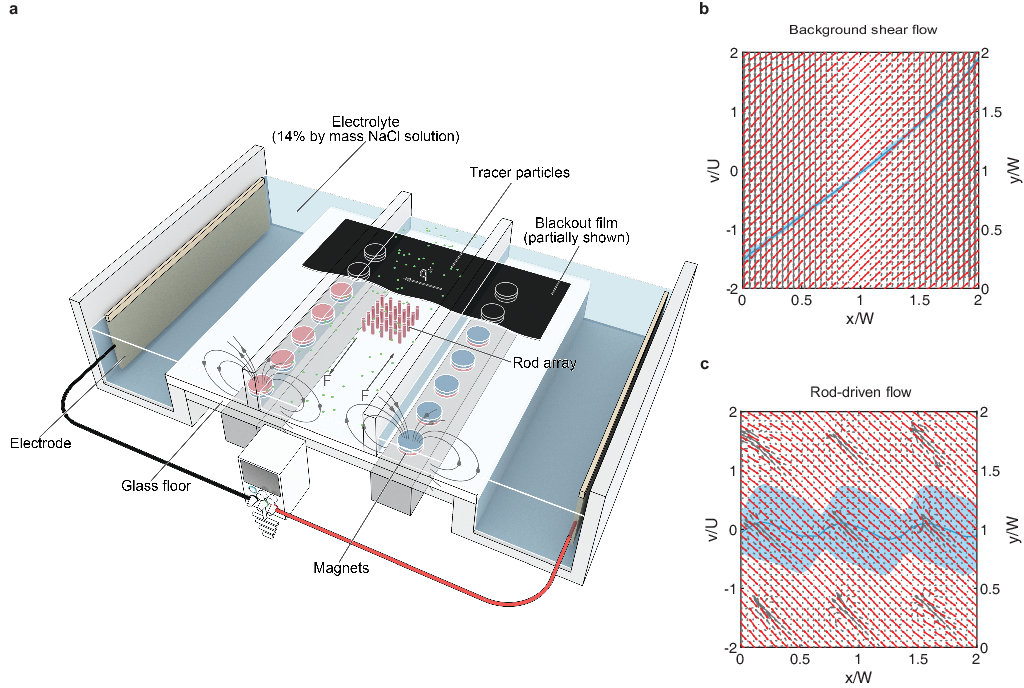} 
\end{figure}
\begin{figure}
    \caption{\textbf{Experimental setup and the characterization of the background flow and physical perturbations} \textbf{a}, A schematic of the experimental setup. The cross-section of the experimental setup illustrates the thin fluid layer and tracer particles at the surface (not to scale). A pair of electrodes conducts direct current horizontally through the electrolyte. The vertical magnetic field from the permanent magnets interacts with the horizontal direct current to generate the Lorentz force on the fluid, which acts nearly within the plane. The tracer particle on the fluid’s surface represents the 2D space under study. A rod array is controlled by a linear actuator, which can introduce directionally biased physical perturbations. 
    \textbf{b}, The flow field of the hydrodynamic shear (gray arrows). Red double-headed arrows indicate the extensional direction of $s_{ij}^{(\rm L)}$ ($\hat{\sigma}$).  \textbf{c}, Flow field of a moving rod array in quiescent fluid (gray arrows). Red double-headed arrows indicate the extensional direction of $\tau_{ij}^{(\rm L)}$ ($\hat{\gamma}$). The blue curves in \textbf{b} and \textbf{c} represent the assembled average of the $v$ component of velocity along the x-axis, normalized by the root-mean-square velocity U. Shaded areas indicate the standard deviation of the $v$ component of velocity normalized by U. The assembled average was calculated both temporally and spatially along the y-axis. Arrows in the velocity and eigenvector fields were down-sampled for clearer visualization.
    }
    \label{fig2} 
\end{figure}

\begin{figure}[p]
    \centering
    \includegraphics[width=\textwidth]{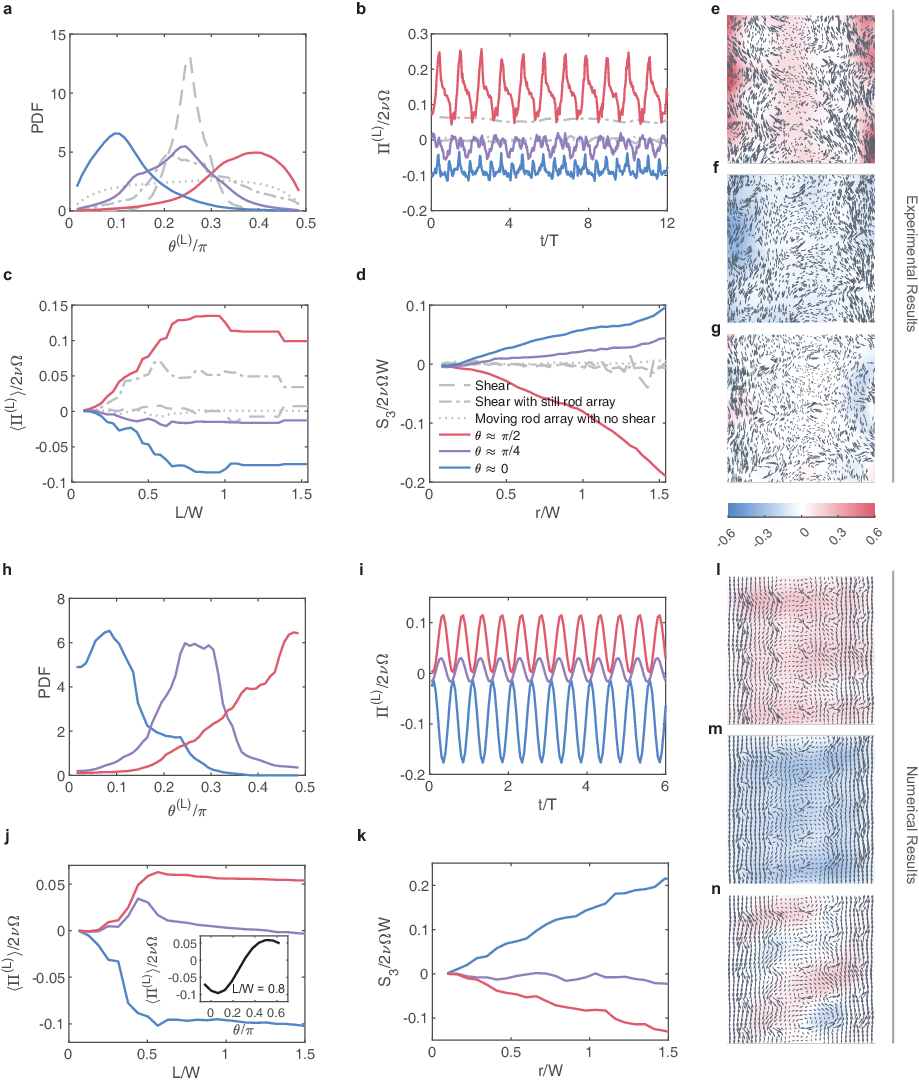} 
\end{figure}
\begin{figure}
    \caption{\textbf{Experimental and numerical results of energy flux manipulation. } \textbf{a} and \textbf{h}, Probability density functions (PDFs) of tensor alignment angle ($\theta^{(\rm L)}$) with L/W = 0.8 for both experiments and simulations, where W is half of the domain size for experiments and simulations. \textbf{b} and \textbf{i}, Temporal evolution of spatially averaged $\Pi^{(\rm L)}$ with L/W = 0.8 for experiments and simulations, respectively. \textbf{c} and \textbf{j}, $\Pi^{(\rm L)}$ at different L for experiments and simulations, respectively. The inset of \textbf{j} is the $\Pi^{(\rm L)}$ at L/W = 0.8 for a range of mechanical angle ($\theta$). \textbf{d} and \textbf{k}, Third-order structure function $S_3$ at different displacement r for experiments and simulations, respectively. \textbf{e-g} and \textbf{l-n}, Snapshots of spatial distribution of energy flux for $\theta \approx \pi/2$, $\theta \approx \pi/4$, and $\theta \approx 0$, respectively.  \textbf{e-g} correspond to experiments and \textbf{l-n} correspond to simulations. All times are normalized by the rod array's moving period for experimental results and by the blinking period of the monopole array for simulation results. All lengths are normalized by half the domain size. $\Pi^{(\rm L)} $ is normalized by viscous dissipation $2\nu \Omega$, where $\Omega$ is the spatially averaged vorticity square. 
    }
    \label{fig3}
\end{figure}

\begin{figure}[p]
    \centering
    \includegraphics[width=\textwidth]{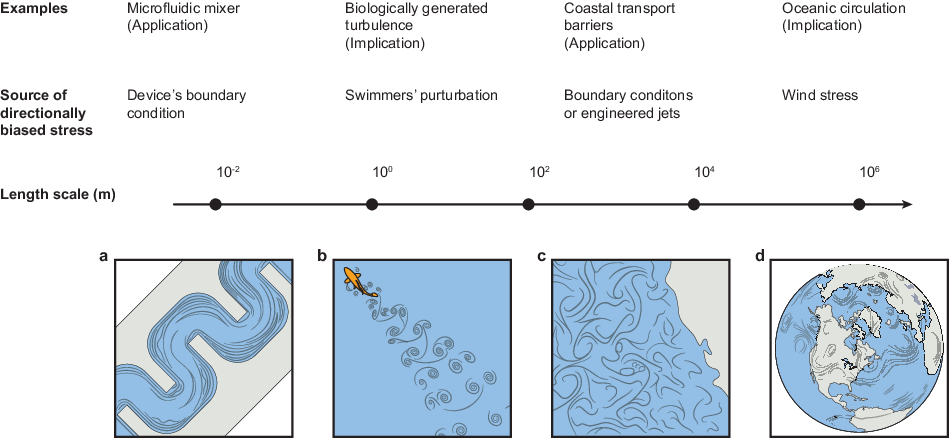} 
    \caption{\textbf{Applications and implications in natural and engineered systems.} Spectral energy flux manipulation can occur in systems spanning scales from $10^{-3}$ to $10^{6}$ meters. This manipulation occurs either through engineering (\textbf{a} and \textbf{c}) or via natural processes. \textbf{a}, By engineering appropriate boundary conditions that interact with fluid to generate directionally biased stress, we can force forward energy flux even at a Reynolds number of order 1, as our theoretical framework remains valid. The increase in small-scale energy will generate fine-scale eddies that facilitate mixing in microfluidic mixers. \textbf{b}, In nature, biologically generated agitation is found to be able to couple with the background hydrodynamic shear to generate either forward or inverse energy flux. This process can either attenuate or strengthen the background hydrodynamic shear, impacting the local biogeochemical structure of the water column. \textbf{c}, Engineered boundary conditions or directionally biased jets with moderate energy can significantly impact LCSs in coastal oceans. In the Methods section, we present a theoretical estimation of the energy power needed to actively manipulate LCSs. \textbf{d}, Climate change will profoundly alter wind fields and oceanic flows. Our results suggest that the altered wind stress could profoundly affect the direction of energy flux in the oceanic flow due to varying alignments between wind stress and oceanic flow. 
    }
    \label{fig4}
\end{figure}


\clearpage 

%
\bibliography{Bib_final.bib} 

\begin{thebibliography}{10}
\providecommand{\url}[1]{\texttt{#1}}
\expandafter\ifx\csname urlstyle\endcsname\relax
  \providecommand{\doi}[1]{doi:\discretionary{}{}{}#1}\else
  \providecommand{\doi}{doi:\discretionary{}{}{}\begingroup
  \urlstyle{rm}\Url}\fi

\bibitem{kolmogorov1941local}
A.~N. Kolmogorov, The local structure of turbulence in incompressible viscous
  fluid for very large Reynolds. \emph{Numbers. In Dokl. Akad. Nauk SSSR}
  \textbf{30}, 301 (1941).

\bibitem{kraichnan1967inertial}
R.~H. Kraichnan, Inertial ranges in two-dimensional turbulence. \emph{The
  Physics of Fluids} \textbf{10}~(7), 1417--1423 (1967).

\bibitem{batchelor1969computation}
G.~K. Batchelor, Computation of the energy spectrum in homogeneous
  two-dimensional turbulence. \emph{The Physics of Fluids} \textbf{12}~(12),
  II--233 (1969).

\bibitem{kolmogorov1962refinement}
A.~N. Kolmogorov, A refinement of previous hypotheses concerning the local
  structure of turbulence in a viscous incompressible fluid at high Reynolds
  number. \emph{Journal of Fluid Mechanics} \textbf{13}~(1), 82--85 (1962).

\bibitem{pope_2000}
S.~B. Pope, \emph{Turbulent Flows} (Cambridge University Press) (2000),
  \doi{10.1017/CBO9781316179475}.

\bibitem{richardson1920supply}
L.~F. Richardson, The supply of energy from and to atmospheric eddies.
  \emph{Proceedings of the Royal Society of London. Series A, Containing Papers
  of a Mathematical and Physical Character} \textbf{97}~(686), 354--373 (1920).

\bibitem{xia2011upscale}
H.~Xia, D.~Byrne, G.~Falkovich, M.~Shats, Upscale energy transfer in thick
  turbulent fluid layers. \emph{Nature Physics} \textbf{7}~(4), 321--324
  (2011).

\bibitem{fang2017multiple}
L.~Fang, N.~T. Ouellette, Multiple stages of decay in two-dimensional
  turbulence. \emph{Physics of Fluids} \textbf{29}~(11), 111105 (2017).

\bibitem{fang2021spectral}
L.~Fang, N.~T. Ouellette, Spectral condensation in laboratory two-dimensional
  turbulence. \emph{Physical Review Fluids} \textbf{6}~(10), 104605 (2021).

\bibitem{fang2016advection}
L.~Fang, N.~T. Ouellette, Advection and the efficiency of spectral energy
  transfer in two-dimensional turbulence. \emph{Physical review letters}
  \textbf{117}~(10), 104501 (2016).

\bibitem{liao2014geometry}
Y.~Liao, N.~T. Ouellette, Geometry of scale-to-scale energy and enstrophy
  transport in two-dimensional flow. \emph{Physics of Fluids} \textbf{26}~(4)
  (2014).

\bibitem{si2024interaction}
X.~Si, L.~Fang, Interaction between swarming active matter and flow: The impact
  on Lagrangian coherent structures. \emph{Physical Review Fluids}
  \textbf{9}~(3), 033101 (2024).

\bibitem{si2024biologically}
X.~Si, L.~Fang, Biologically generated turbulent energy flux in shear flow
  depends on tensor geometry. \emph{PNAS nexus} \textbf{3}~(2), pgae056 (2024).

\bibitem{ouellette2006quantitative}
N.~T. Ouellette, H.~Xu, E.~Bodenschatz, A quantitative study of
  three-dimensional Lagrangian particle tracking algorithms. \emph{Experiments
  in Fluids} \textbf{40}, 301--313 (2006).

\bibitem{boyd2001chebyshev}
J.~P. Boyd, \emph{Chebyshev and Fourier spectral methods} (Courier Corporation)
  (2001).

\bibitem{valadao2024spectrum}
V.~J. Valadão, G.~Boffetta, M.~Crialesi-Esposito, F.~D. Lillo, S.~Musacchio,
  Spectrum correction on {Ekman-Navier-Stokes} equation in two-dimensions
  (2024), \url{https://arxiv.org/abs/2408.15735}.

\bibitem{kolmogorov1991dissipation}
A.~N. Kolmogorov, Dissipation of energy in the locally isotropic turbulence.
  \emph{Proceedings of the Royal Society of London. Series A: Mathematical and
  Physical Sciences} \textbf{434}~(1890), 15--17 (1991).

\bibitem{alexakis2018cascades}
A.~Alexakis, L.~Biferale, Cascades and transitions in turbulent flows.
  \emph{Physics Reports} \textbf{767}, 1--101 (2018).

\bibitem{stroock2002chaotic}
A.~D. Stroock, \emph{et~al.}, Chaotic mixer for microchannels. \emph{Science}
  \textbf{295}~(5555), 647--651 (2002).

\bibitem{stone2004engineering}
H.~A. Stone, A.~D. Stroock, A.~Ajdari, Engineering flows in small devices:
  microfluidics toward a lab-on-a-chip. \emph{Annu. Rev. Fluid Mech.}
  \textbf{36}~(1), 381--411 (2004).

\bibitem{houghton2018vertically}
I.~A. Houghton, J.~R. Koseff, S.~G. Monismith, J.~O. Dabiri, Vertically
  migrating swimmers generate aggregation-scale eddies in a stratified column.
  \emph{Nature} \textbf{556}~(7702), 497--500 (2018).

\bibitem{kunze2006observations}
E.~Kunze, J.~F. Dower, I.~Beveridge, R.~Dewey, K.~P. Bartlett, Observations of
  biologically generated turbulence in a coastal inlet. \emph{Science}
  \textbf{313}~(5794), 1768--1770 (2006).

\bibitem{katija2012biogenic}
K.~Katija, Biogenic inputs to ocean mixing. \emph{Journal of Experimental
  Biology} \textbf{215}~(6), 1040--1049 (2012).

\bibitem{olascoaga2006persistent}
M.~Olascoaga, \emph{et~al.}, Persistent transport barrier on the West Florida
  Shelf. \emph{Geophysical research letters} \textbf{33}~(22) (2006).

\bibitem{fang2018influence}
L.~Fang, N.~T. Ouellette, Influence of lateral boundaries on transport in
  quasi-two-dimensional flow. \emph{Chaos: An Interdisciplinary Journal of
  Nonlinear Science} \textbf{28}~(2), 023113 (2018).

\bibitem{dewar1987some}
W.~K. Dewar, G.~R. Flierl, Some effects of the wind on rings. \emph{Journal of
  physical oceanography} \textbf{17}~(10), 1653--1667 (1987).

\bibitem{renault2016control}
L.~Renault, M.~J. Molemaker, J.~Gula, S.~Masson, J.~C. McWilliams, Control and
  stabilization of the Gulf Stream by oceanic current interaction with the
  atmosphere. \emph{Journal of Physical Oceanography} \textbf{46}~(11),
  3439--3453 (2016).

\bibitem{rai2021scale}
S.~Rai, M.~Hecht, M.~Maltrud, H.~Aluie, Scale of oceanic eddy killing by wind
  from global satellite observations. \emph{Science Advances} \textbf{7}~(28),
  eabf4920 (2021).

\bibitem{germano1992turbulence}
M.~Germano, Turbulence: the filtering approach. \emph{Journal of Fluid
  Mechanics} \textbf{238}, 325--336 (1992).

\bibitem{rivera2003energy}
M.~Rivera, W.~Daniel, S.~Chen, R.~Ecke, Energy and enstrophy transfer in
  decaying two-dimensional turbulence. \emph{Physical review letters}
  \textbf{90}~(10), 104502 (2003).

\bibitem{leonard1975energy}
A.~Leonard, Energy cascade in large-eddy simulations of turbulent fluid flows,
  in \emph{Advances in geophysics} (Elsevier), vol.~18, pp. 237--248 (1975).

\bibitem{liao2013spatial}
Y.~Liao, N.~T. Ouellette, Spatial structure of spectral transport in
  two-dimensional flow. \emph{Journal of Fluid Mechanics} \textbf{725},
  281--298 (2013).

\bibitem{ballouz2018tensor}
J.~G. Ballouz, N.~T. Ouellette, Tensor geometry in the turbulent cascade.
  \emph{Journal of Fluid Mechanics} \textbf{835}, 1048--1064 (2018).

\bibitem{xiao2009physical}
Z.~Xiao, M.~Wan, S.~Chen, G.~Eyink, Physical mechanism of the inverse energy
  cascade of two-dimensional turbulence: a numerical investigation.
  \emph{Journal of Fluid Mechanics} \textbf{619}, 1--44 (2009).

\bibitem{ouellette2008transport}
N.~T. Ouellette, P.~O’Malley, J.~P. Gollub, Transport of finite-sized
  particles in chaotic flow. \emph{Physical review letters} \textbf{101}~(17),
  174504 (2008).

\bibitem{vella2005cheerios}
D.~Vella, L.~Mahadevan, The “cheerios effect”. \emph{American journal of
  physics} \textbf{73}~(9), 817--825 (2005).

\bibitem{schneider2005decaying}
K.~Schneider, M.~Farge, Decaying two-dimensional turbulence in a circular
  container. \emph{Physical review letters} \textbf{95}~(24), 244502 (2005).

\end{thebibliography}
\bibliographystyle{sciencemag}

%
%
%
%
%
%


\paragraph*{Funding:}
L.F. thanks the U.S. National Science Foundation under Grant No. \mbox{CMMI-2143807} and \mbox{CBET-2429374}.
\paragraph*{Author contributions:}
L.F. conceived the original idea and supervised the project. X.S. ran the
experiments and analyzed the data. G.B. and F.D.L. ran the simulation. All authors wrote the paper.
\paragraph*{Competing interests:}
There are no competing interests to declare.
\subsection*{Supplementary materials}
Materials and Methods\\
Figs. S1 to S4\\
References \textit{(7-\arabic{enumiv})} 


\newpage


\renewcommand{\thefigure}{S\arabic{figure}}
\renewcommand{\thetable}{S\arabic{table}}
\renewcommand{\theequation}{S\arabic{equation}}
\renewcommand{\thepage}{S\arabic{page}}
\setcounter{figure}{0}
\setcounter{table}{0}
\setcounter{equation}{0}
\setcounter{page}{1} 


\begin{center}
\section*{Supplementary Materials for\\ \scititle}

Xinyu~Si$^{1}$,
Filippo~De~Lillo$^{3}$,
Guido~Boffetta$^{3}$
Lei~Fang$^{1,2\ast}$\\
\small$^\ast$Corresponding author. Email: lei.fang@pitt.edu
\end{center}

\subsubsection*{This PDF file includes:}
Materials and Methods\\
Figures S1 to S4

\newpage


\subsection*{Materials and Methods}

\subsection{The filtering approach and the spectral energy flux term}

\subsubsection{Filter space technique (FST)}
The filter space technique (FST) is based on a filtering process \cite{germano1992turbulence} and can extract spatially localized scale-to-scale energy flux information from measured flow fields \cite{rivera2003energy}.
The filtering process can be generally expressed as a convolutional integral \cite{leonard1975energy}. For example, the filtered component of a velocity field has the form
\begin{equation}\label{Eqn_Fitering}
    u_i^{(\rm L)} (\mathbf{x}) \equiv \int G^{(\rm L)}(\mathbf{r},\mathbf{x}) u_i(\mathbf{x} - \mathbf{r})d\mathbf{r},
\end{equation}
where $G^{(\rm L)}$ is a kernel acting as a low-pass filter, with the superscript L indicating the cut-off length scale. Our result is not sensitive to the specific nature of the filter kernel. In this paper, we used a sharp spectral filter (with a cutoff length L) smoothed by a Gaussian window to avoid the ringing effect.

To obtain the spectral energy flux term $\Pi^{(\rm L)}$, we start from filtering the Navier-Stokes equations that govern the motion for incompressible fluids:
\begin{equation}\label{Eqn_2DNS}
    \frac{\partial u_i}{\partial {\rm t}} + u_j \frac{\partial u_i}{\partial x_j} = -\frac{1}{\rho}\frac{\partial p}{\partial x_i} + \nu\frac{\partial^2 u_i}{\partial x_j \partial x_j} \quad \mbox{and} \quad \frac{\partial u_i}{\partial x_i} = 0,
\end{equation}
in which $u_i$ is the i$^{th}$ component of velocity, $\rho$ the density, $p$ the pressure and $\nu$ the kinematic viscosity. After applying the filter, we can obtain the evolution equation for the filtered velocity field $u_i^{(\rm L)}$ as: 
\begin{equation}\label{Eqn_Filtered_2DNS}
    \frac{\partial u_i^{(\rm L)}}{\partial {\rm t}} + u_j^{(\rm L)} \frac{\partial u_i^{(\rm L)}}{\partial x_j} = -\frac{1}{\rho} \frac{\partial p^{(\rm L)}}{\partial x_i} + \nu\frac{\partial^2 u_i^{(\rm L)}}{\partial x_j \partial x_j} - \frac{\partial \tau_{ij}^{(\rm L)}}{\partial x_j},
\end{equation}
where $\tau_{ij}^{(\rm L)} = (u_iu_j)^{(\rm L)} - u_i^{(\rm L)}u_j^{(\rm L)}$.

Taking the inner product of $u_i^{(\rm L)}$ and the filtered momentum equation \ref{Eqn_Filtered_2DNS}, we can obtain the equation of motion for the filtered kinetic energy ${\rm E}^{(\rm L)} = \frac{1}{2}u_i^{(\rm L)}u_i^{(\rm L)}$ as
\begin{equation}\label{Eqn_MotionEnergy}
    \frac{\partial {\rm E}^{(\rm L)}}{\partial {\rm t}} = - \frac{\partial J_i^{(\rm L)}}{\partial x_i} - \nu \frac{\partial u_i^{(\rm L)}}{\partial x_j}\frac{\partial u_j^{(\rm L)}}{\partial x_i} - \Pi^{(\rm L)},
\end{equation}
where $\Pi^{(\rm L)} = -\tau_{ij}^{(\rm L)}s_{ij}^{(\rm L)}$ with $s_{ij}^{(\rm L)} = \frac{1}{2}(\partial u_i^{(\rm L)}/\partial x_j + \partial u_j^{(\rm L)}/\partial x_i)$ the rate of strain tensor for the filtered velocity field. 
On the right-hand side of Eqn. \ref{Eqn_MotionEnergy}, the term with $J_{i}^{(\rm L)}$ assembles all the terms that represents the spatial currents of filtered energy. The second term represents the viscous damping of energy within the resolved scales. Term $\Pi^{(\rm L)}$, in particular, represents the spectral energy flux between scales smaller than L and scales larger than L. $\Pi^{(\rm L)} < 0$ indicates inverse energy flux toward larger length scales. $\Pi^{(\rm L)} > 0$ indicates forward energy flux toward smaller length scales.

\subsubsection{Spectral energy flux term decomposition}
The stress tensor $\tau_{ij}^{(\rm L)}$ can be further decomposed into three components \cite{leonard1975energy, pope_2000, liao2013spatial} based on the type of triad interactions as:
\begin{eqnarray}\label{Eqn_TauDecomp}
    \tau_{ij}^{(\rm L)} & = & {(u_i^{(\rm L)}u_j^{(\rm L)})}^{(\rm L)} - u_i^{(\rm L)}u_j^{(\rm L)} \nonumber \\
    & + & {(u_i^{(\rm L)}(u_j - u_j^{(\rm L)}))}^{(\rm L)} + {(u_j^{(\rm L)}(u_i - u_i^{(\rm L)}))}^{(\rm L)} \nonumber\\
    & + & {((u_i - u_i^{(\rm L)})(u_j - u_j^{(\rm L)}))}^{(\rm L)}.
\end{eqnarray}
The first component $\tau_L^{(\rm L)} = {(u_i^{(\rm L)}u_j^{(\rm L)})}^{(\rm L)} - u_i^{(\rm L)}u_j^{(\rm L)}$ is a small-scale quantity composed of two large-scale quantities.
The second component $\tau_C^{(\rm L)} = {(u_i^{(\rm L)}(u_j - u_j^{(\rm L)}))}^{(\rm L)} + {(u_j^{(\rm L)}(u_i - u_i^{(\rm L)}))}^{(\rm L)}$ is a large-scale quantity composed of one large-scale quantity and one small-scale quantity. 
The third component $\tau_S^{(\rm L)} = {((u_i - u_i^{(\rm L)})(u_j - u_j^{(\rm L)}))}^{(\rm L)}$ is a large-scale quantity composed of two small-scale quantities. 
In large eddy simulation (LES), the three terms are called Leonard stress, cross stress, and subgrid-scale Reynolds stress respectively. We take these names in this paper for convenience. However, note that $u_i$ here contains information about all scales of motion and hence the term $\tau_{ij}^{(\rm L)}$ involves no modeling, which is different from LES.
The inner products of these components with $s_{ij}^{(\rm L)}$ give the corresponding components of the spectral energy flux $\Pi^{(\rm L)}$ as $\Pi_L^{(\rm L)}$, $\Pi_C^{(\rm L)}$ and $\Pi_S^{(\rm L)}$:
\begin{eqnarray}\label{Eqn_PiDecomp}
    \Pi^{(\rm L)} & = &\Pi_L^{(\rm L)} + \Pi_C^{(\rm L)} + \Pi_S^{(\rm L)} \\
    \Pi_L^{(\rm L)} & = & - \tau_L^{(\rm L)}s_{ij}^{(\rm L)} \\
    \Pi_C^{(\rm L)} & = & - \tau_C^{(\rm L)}s_{ij}^{(\rm L)} \\
    \Pi_S^{(\rm L)} & = & - \tau_S^{(\rm L)}s_{ij}^{(\rm L)}.
\end{eqnarray}

In \cite{liao2013spatial}, it has been shown that the subgrid term $\Pi_S^{(\rm L)}$ carries most of the net spectral energy flux information between the large and small scales. The Leonard term and the cross term involve more subtle interpretations. Using a simple cellular flow, \cite{liao2013spatial} showed that the Leonard term is dominated by the transfer of energy between different resolved wavevectors rather than the transfer of energy between large and small scales. However, after spatial averaging, they found that the Leonard term and crossing term have a negligible contribution to the net spectral energy flux, which is also verified by our experiment data of a 2D turbulent flow (see Extended Data Fig. 1).

Despite the theoretically negligible contribution to the spectral energy flux by the Leonard term and the cross term, including the Leonard term and the cross term will cause contamination to the calculated spatially averaged spectral energy flux, especially in regions near boundaries. Specifically, this contamination comes from edge padding when applying a filter to the measured data near boundaries. Padding involves filling artificial data (in this paper, we used zero-padding) into the regions out of boundaries where there are no measured data so that the filtering process can be applied near boundaries. The magnitude of this padding error is small compared to the magnitude of the small-scale fluctuations ($u_i - u_i^{(\rm L)}$) created by the moving rods. However, this padding error will be magnified when it is added to or multiplied by a large-scale velocity ($u_i^{(\rm L)}$).
Therefore, in this paper, we used the subgrid flux term $\Pi_S^{(\rm L)}$ in our analysis instead of $\Pi^{(\rm L)}$. For simplicity and clarity, we omitted the subscript in both the figures and the main text.



\subsection{Theoretical background for tensor geometry}

\subsubsection{Rewriting the spectral energy flux term}
The theory of tensor geometry is described in detail in \cite{fang2016advection,ballouz2018tensor}. Here, we just briefly introduce the necessary information. Since the rate of strain tensor $s_{ij}^{(\rm L)}$ is symmetric and deviatoric, we can just consider the deviatoric part of the stress tensor when calculating the spectral energy flux because only this part will impact the inner product \cite{xiao2009physical}. 
Both the rate of strain tensor and the deviatoric part of the stress tensor have two eigenvalues with the same magnitude and opposite sign. The eigenvector corresponding to the positive eigenvalue (referred to as extensional) and the eigenvector corresponding to the negative eigenvalue (referred to as compressional) are orthogonal. We label the extensional eigenvalue for the stress tensor as $\gamma$ and that for the rate of strain tensor as $\sigma$. Working in the eigenbasis of the stress tensor and marking the angle between the extensional eigenvectors of these two tensors as $\theta^{(\rm L)}$, we have
\begin{eqnarray}\label{Eqn_TG}
    -\Pi^{(\rm L)} & = & \tau_{ij}^{(\rm L)}s_{ij}^{(\rm L)} \nonumber \\
    & = & \mbox{Tr} \left[
    \left(\begin{array}{cc}
         \gamma & 0 \\
         0 & \gamma
    \end{array}\right) 
    \left(\begin{array}{cc}
         \cos (\theta^{(\rm L)}) & -\sin(\theta^{(\rm L)})  \\
         \sin(\theta^{(\rm L)}) & \cos (\theta^{(\rm L)})
    \end{array} \right)
    \left(\begin{array}{cc}
         \sigma & 0 \\
         0 & \sigma
    \end{array}\right)
    \left(\begin{array}{cc}
         \cos (\theta^{(\rm L)}) & \sin(\theta^{(\rm L)})  \\
         -\sin(\theta^{(\rm L)}) & \cos (\theta^{(\rm L)})
    \end{array} \right)
    \right] \nonumber \\
    & = & 2\gamma\sigma\cos(2\theta^{(\rm L)}).
\end{eqnarray}
Note that using $\tau_{ij}^{(\rm L)}$ or $\tau_S^{(\rm L)}$ does not affect the derivation of Eqn. \ref{Eqn_TG}.
Since both $\gamma$ and $\sigma$ are positive, we can see that the direction of spectral energy flux depends only on the alignment of the eigenframes of the two tensors.

\subsubsection{Tensor geometry of the large-scale shear}
Through the perspective of tensor geometry, there arises the possibility of manipulating the spectral energy flux by forcing the small-scale stress to align with the large-scale rate of strain in any intended angles. 
To demonstrate this, first consider a cut-off length scale L. At large scales, there exists a steady shear flow whose width is much larger than L. 
To simplify this problem, we set the steady shear flow with streamlines aligning with the y axis and with no stream-wise velocity gradient. In the x direction, the shear has a constant velocity gradient K for the vertical velocity component. The rate of strain tensor of the shear flow at any length scale L is then
\begin{equation}\label{Eqn_Sij}
    s_{ij}^{(\rm L)} = 
    \begin{bmatrix}
        0 & \frac{1}{2} \rm K \\ 
        \frac{1}{2} \rm K & 0
    \end{bmatrix}
    .
\end{equation}
Since this matrix is traceless and symmetric, it has two eigenvalues of the same magnitude but with opposite signs. The two eigenvectors are orthogonal and the angle between the extensional eigenvector and the x-axis has an angle of $\pi/4$ (or 5$\pi$/4). 

If we apply disturbances to the large-scale shear flow with injection length scales much smaller than $L$, the nonlinear coupling between the applied small-scale stresses and the background flow will result in turbulent flow that transfers energy through scales. 

\subsubsection{Tensor geometry of the small-scale stresses}
Here, we demonstrate how we generate engineered small-scale stress through physical perturbations. For simplicity, consider the small-scale disturbance as a velocity vector $b_i$ that forms an angle $\theta_b$ with the x-axis.
Given enough scale separation between the small-scale disturbance and the large-scale shear, we would expect that most of the information induced by the small-scale disturbance will be included in the residue after filtering. 
Therefore, we can estimate that $b_i \approx u_i - u_i^{(\rm L)}$. The deviatoric part of the subgrid-scale Reynolds stress is thus
\begin{eqnarray}\label{Eqn_Stress}
    \tau_S^{(\rm L)} & = & (b_ib_j)^{(\rm L)} \nonumber \\
    & = & \left( \left\|b_i\right\|^2 \begin{bmatrix}
    \cos^2(\theta_b) - \frac{1}{2} & \sin(\theta_b)*\cos(\theta_b)  \\
    \sin(\theta_b)*\cos(\theta_b) & \sin^2(\theta_b) - \frac{1}{2} 
    \end{bmatrix} \right) ^{(\rm L)}.
\end{eqnarray}
We note that the filtering process will not affect the eigenvector direction. We can get that the extensional eigenvector for the deviatoric subgrid-scale Reynolds stress $\tau_S^{(\rm L)}$ is in the direction of $\begin{pmatrix} 1  \\ tan(\theta_{b}) \end{pmatrix}$. Therefore, the direction of the extensional eigenvector of $\tau_S^{(\rm L)}$ is in parallel with the direction of $b_i$.

From the derivations above, we can see that the directions of the extensional eigenvectors for both $s_{ij}^{(\rm L)}$ and $\tau_S^{(\rm L)}$ are known even before applying physical perturbations to the background flows. Therefore, based on Eqn. \ref{Eqn_TG}, it is possible to manipulate the direction of spectral energy flux by controlling and alignment between the eigenframe of these two tensors. 

\subsection{Quasi-2D turbulence experiments}

\subsubsection{Apparatus and particle tracking}
The main body for the quasi-2D flow system consisted of an acrylic frame, a pair of copper electrodes installed on the opposite sides of the setup, and a piece of tempered glass in the center separating a thin layer of salt water on top and an array of cylindrical magnets below.
The dimensions of the main frame and the glass floor in the center were 96.5 $\times$ 83.8 cm$^2$ and 81.3 $\times$ 81.3 cm$^2$, respectively. The upper surface of the glass was coated with hydrophobic materials (Rain-X) to reduce friction, and the lower surface was covered by light-absorbing blackout film. 
Beneath the glass, cylindrical magnets were organized in desired patterns to drive flow in different directions. Each magnet (neodymium grade N52) had an outer diameter of 1.27 cm and thickness of 0.64 cm, with the maximum magnetic flux density of 1.5 T at the magnet surface. We loaded a thin layer (6 mm thickness) of 14$\%$ by mass NaCl solution on top of the glass. The solution had density $\rho =$ 1.101g/cm$^3$ and viscosity $\nu = 1.25 \times 10^{-2}$ cm$^2$/s. 
By passing a DC current through the conducting solution layer, we were able to drive a quasi-2D flow with the resulting Lorentz body force and control the flow Reynolds number by adjusting the DC current intensity. The 2D was well-kept throughout our experiments.

To track the flow, we seeded green fluorescent polyethylene tracer particles (Cospheric) into the fluid. The tracer particles had a density of 1.025 g/cm$^3$ and diameters ranging from 106 to 125 $\mu$m. The Stokes number of the particles was of order 10$^{-3}$, which means that the particle could accurately trace the flow \cite{ouellette2008transport}. Since the density of the particles was lower than the working fluid, they would float on the gas-liquid interface. Due to surface tension effects, they would show a slow clustering tendency, which is known as the ``cheerios effect" \cite{vella2005cheerios}. To reduce the surface tension, a small amount of surfactant was added to the fluid in order to minimize the impact on tracer movements. Our measurement of tracers in quiescent fluid showed that the ``cheerios effect" was negligible.

We used a machine vision camera (Basler, acA2040-90$\mu$m) to image the flow that was illuminated by blue LED lights. An 11.4 cm by 11.4 cm region at the center of the setup was recorded with a resolution of 1,600 pixels by 1,600 pixels. About 18,000 to 22,000 particles could be recorded at a frame rate of 60 frames per second. With this particle density and frame rate, we could obtain highly spatiotemporally resolved velocity fields through a particle tracking velocimetry (PTV) algorithm \cite{ouellette2006quantitative}. For easier use, the measured flow was then interpolated onto regular Eulerian grids using cubic interpolation with a grid size of 12 pixels (0.85 mm), which gave a grid density not higher than the original particle density. The final analysis to obtain Fig. \ref{fig2} and \ref{fig3} was performed on a 7.4 cm by 7.4 cm domain at the center of the measured area to reduce errors near boundaries caused by edge padding during filtering .

\subsubsection{Two-dimensional steady shear flow with moving rods}
We used two stripes of magnets with opposite polarity. The distance between the two stripes was 20 cm. When a dc current was conducted through the fluid, the two stripes generated a hydrodynamic shear with an ordered rate of strain (Fig. \ref{fig2}b). To apply the small-scale stress to couple with the rate of strain in the background flow, we built a 5 by 5 grid of rods and drove the grid with a programmable linear actuator. The diameter of each rod was 2 mm, and the center-to-center space between neighbor rods was 2.5 cm. The rod array moved back and forth at a speed of 1 cm/s to generate directionally biased stress (Fig. \ref{fig2}c). We define Reynolds number Re = UW/$\nu$ where U is the root-mean-square velocity, W is half of the domain width for analysis and $\nu$ is the kinematic viscosity. The Reynolds number of the resulting flow was 210.

\subsection{2D turbulence simulation}
The numerical simulations were carried out using a standard fully dealiased pseudospectral code \cite{boyd2001chebyshev,valadao2024spectrum}. Eqn. ~\ref{Eqn_2DNS} were integrated on a 2D domain of size $L_x \times L_y$, with a second order Runge-Kutta temporal scheme. In order to simulate a configuration similar to that of the experiment we used as a base flow, a linear shear flow was obtained by imposing the boundary conditions $\mathbf{u}(L_x/2,y)=\mathbf{u}_+=(0,+U_s)$ and $\mathbf{u}(-L_x/2,y)=\mathbf{u}_-=(0,-U_s)$ at the walls $x=\pm L_x/2$, with periodic boundary condition on the $y$ direction. The boundary conditions were implemented via a penalization method \cite{schneider2005decaying}. Specifically, at each time step the body force
\begin{equation}
\mathbf{F}_{\rm pen}(\mathbf{x})=-\lambda (\mathbf{u}(\mathbf{x})-\mathbf{u}_{\pm})\phi(x \mp L_x/2)
\label{eqappnum1}
\end{equation}
was imposed, with $\lambda$ a large parameter. The scalar function $\phi$ is a mask with support only within a small distance $r_b$ of the boundaries and defined as $\phi(x)=\cos(\frac{\pi x}{2r_b})$ if $|x|<r_b$ and $\phi(x)=0$ otherwise. All the simulations presented in this paper are performed with shear velocity $U_s=1$ and domain sizes  $L_x=6.136$, $L_y=2\pi$ (arbitrary units). We used a numerical resolution of $N_x\times N_y=500 \times 512$ grid points which is sufficient to resolve the smallest scale in the flow
and the support of each penalization mask was $2r_b=9.8\times10^{-2}$ corresponding to 8 grid points.

The local forcing was applied using 25 force monopoles whose centers were organized on a regular $5\times5$ square grid with side $2.0$ (approximately one-third of the span of the effective numerical channel). Each monopole applied a pulsating force $\mathbf{F}_i=f\sin(\omega t)G(\mathbf{x}_i)\mathbf{e}(\theta_m)$, where $\mathbf{x}_i$ is the position of the $i-$th monopole, $G(\mathbf{x})$ a two-dimensional normalized Gaussian with a half-width of 4 grid points and $\mathbf{e}(\theta_m)=(-\sin\theta_m,\cos\theta_m)$ sets the direction of the force monopole at an angle $\theta_m$ with respect to the $y$ axis. The amplitude of the monopole was $f=0.38$.

All the numerical simulations were started from a fluid at rest $\mathbf{u}=0$ and carried on until a shear flow $\mathbf{u}=(0,2 x U_s/L_x)$ was produced. A kinematic viscosity $\nu=10^{-2}$ was used which corresponds to a Reynolds number ${\rm Re}={\rm U}L_x/(2\nu)=184$ based on half channel width and the root-mean-square velocity $\rm U$. After a steady state is reached, the forcing is applied with $f=0.38$ and $\omega=2\pi/5$. Such parameters were chosen to provide close to maximum effect measured in terms of $\Pi^{(\rm L)}_S/(2\nu\Omega)$, and they were kept fixed for all simulations while changing the value of $\theta_m$. In all cases examined in this paper the resulting flow is periodic with the same periodicity of the local forcing (see main text). The analysis was therefore performed over one period with the same code used for the experimental results. The final analysis was conducted in a 2 by 2 domain at the center of the simulation domain to obtain the results in Fig. \ref{fig3}, where local forcing was actively applied.
For more intense forcing, non-periodic (chaotic or turbulent) flows were observed, as well as solutions where periods longer than the pulsating period of the monopoles appeared. However, no such cases are presented here and they may be the object of future investigations.


\subsection{Theoretical estimation of manipulating LCSs}
To demonstrate the manipulation of hyperbolic LCSs, we consider the Taylor-Green (TG) flow with a period of $\pi$:
\begin{eqnarray}\label{Eqn_TaylorGreen}
    u & = & - \cos(x)\sin(y)  \nonumber \\
    v & = & \sin(x)\cos(y).    
\end{eqnarray}
This flow consists of an array of counter-rotating vortices that are separated by hyperbolic LCSs emanating from hyperbolic points. These hyperbolic LCSs act as transport barriers.
Analyzing the large-scale rate of strain tensor of the TG flow gives the extensional eigenvalue as $\sigma = \sin(x)\sin(y)$, from which it can be clearly seen that the largest $\sigma$ locates at hyperbolic points. The corresponding extensional eigenvector $\hat{\sigma}$ aligns with the direction of the attracting LCSs near hyperbolic points.

We then consider applying directionally biased small-scale disturbance $b_i$ to the TG flow. As previously discussed, the extensional eigenvalue of the disturbing flow can be estimated as $\gamma = \|b_i\|^2/2$ and the direction of $\hat{\gamma}$ aligns with $b_i$. Therefore, the angle between $\hat{\sigma}$ and $\hat{\gamma}$ can be esitimated by the angle $\theta$ between $\hat{\sigma}$ and $b_i$.
The energy flux generated at the locations where the fluctuations are applied can then be estimated as:
\begin{eqnarray}
     \Pi^{(\rm L)} & = & -2\gamma\sigma\cos(2\theta) \nonumber \\
     & = & -\|b_i\|^2\sin(x)\sin(y)\cos(2\theta).
\end{eqnarray}

The strategy to manipulate LCSs is to use directionally biased small-scale physical perturbations to couple with TG flow around hyperbolic points to generate either forward or inverse energy flux to deplete or replenish large-scale energy that sustains the large-scale LCSs. Here, our estimation focuses on weakening LCSs. If our physical perturbation leads to a spatially integrated forward energy flux (${\rm Q_T}$) that is comparable to the power sustaining the local background TG flow ($\rm P_{b,A}$), hypothetically, the local TG flow, as well as the local LCSs, will be weakened because the power that sustains the LCSs is dumped to small scales and dissipated. 


For instance, we consider applying directionally biased small-scale perturbations $b_i$ in a $\pi/5$ by $\pi/5$ square region A centered at the hyperbolic point that forms an angle $\theta = \pi/2$ with $\hat{\sigma}$.
In 2D flow, the energy is mainly dissipated by large-scale friction, and the damping rate can be calculated as $\rm P = \alpha \|\mathbf{u}\|^2$, where $\alpha$ is the frictional damping coefficient. Taking $\alpha = 0.1$ and $b_i = 0.08$, we can calculate the power input required to sustain the large-scale TG flow in a $\pi$ by $\pi$ periodic square to be ${\rm P_b} = \int_0^{\pi}\int_0^{\pi} \alpha (u^2 + v^2) dxdy = 0.49$ and the power required in the $\pi/5$ by $\pi/5$ square to be ${\rm P_{b,A}} = \int_{\rm A} \alpha (u^2+v^2) d {\rm A} = 0.0025$. The power required to sustain the small-scale fluctuation in region A should be on the order of ${\rm P_f} = \alpha \|b_i\|^2\pi^2/25 = 2.5\times 10^{-4}$. Integrating the energy flux in the manipulated square region, we have the energy flux as ${\rm Q_T} = \int_{\rm A} \Pi^{(\rm L)} d {\rm A} = \int_{\rm A} \|b_i\|^2\sin(x)\sin(y) d {\rm A} = 0.382\|b_i\|^2 = 0.0024$. With ${\rm Q_T}$ having the similar intensity to ${\rm P_{b,A}}$, we would expect the LCSs to be significantly weakened near the hyperbolic points. Remarkably, the power input to generate the small-scale fluctuation is only 0.05$\%$ of the total energy for sustaining the TG in a $\pi$ by $\pi$ periodic square. This estimation shows that we can use a small amount of power to manipulate large-scale geophysical LCSs to enhance coastal ecosystem health.


\begin{figure}[p]
    \centering
    \includegraphics[width=\textwidth]{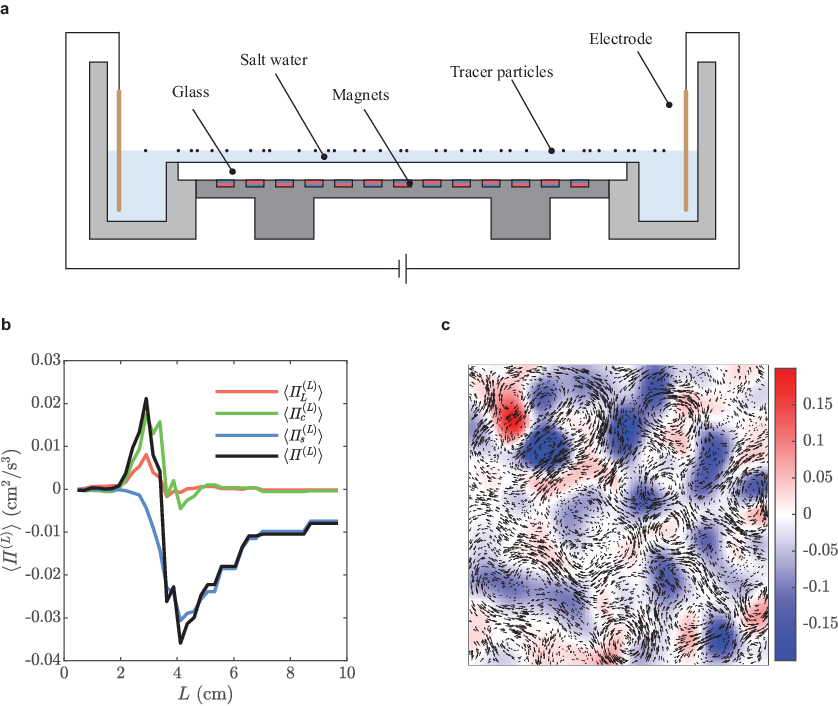}
    \caption{\textbf{Quasi-2D turbulence and spectral energy flux decomposition.} \textbf{a}, Schematic diagram of the experimental setup to generate weak quasi-2D turbulence. This setup used a magnet array with checkerboard pattern of alternating polarities. The center-to-center distance $L_{m}$ between neighbor magnets was 2.54 cm. \textbf{b} shows the spatially and temporally averaged spectral energy flux $\langle\Pi^{({\rm L})}\rangle$ and its components. It can be observed that at scales larger than $L_m$, the net spectral energy flux is inverse and is dominated by the subgrid term ($\Pi_S^{({\rm L})}$) in energy flux range, i.e., length scales greater than the energy injection scale. The Leonard term and the cross term only carry minor information. \textbf{c} shows a snapshot of the $\Pi_S^{({\rm L})}$ map for the measured weak turbulent flow field with the cutoff length $L$ = 4 cm. It can be noticed that even though the spatially averaged spectral energy flux is inverse, both inverse and forward energy flux can occur.
    }
    \label{Ext.fig1}
\end{figure}

\begin{figure}[p]
    \centering
    \includegraphics[width=\textwidth]{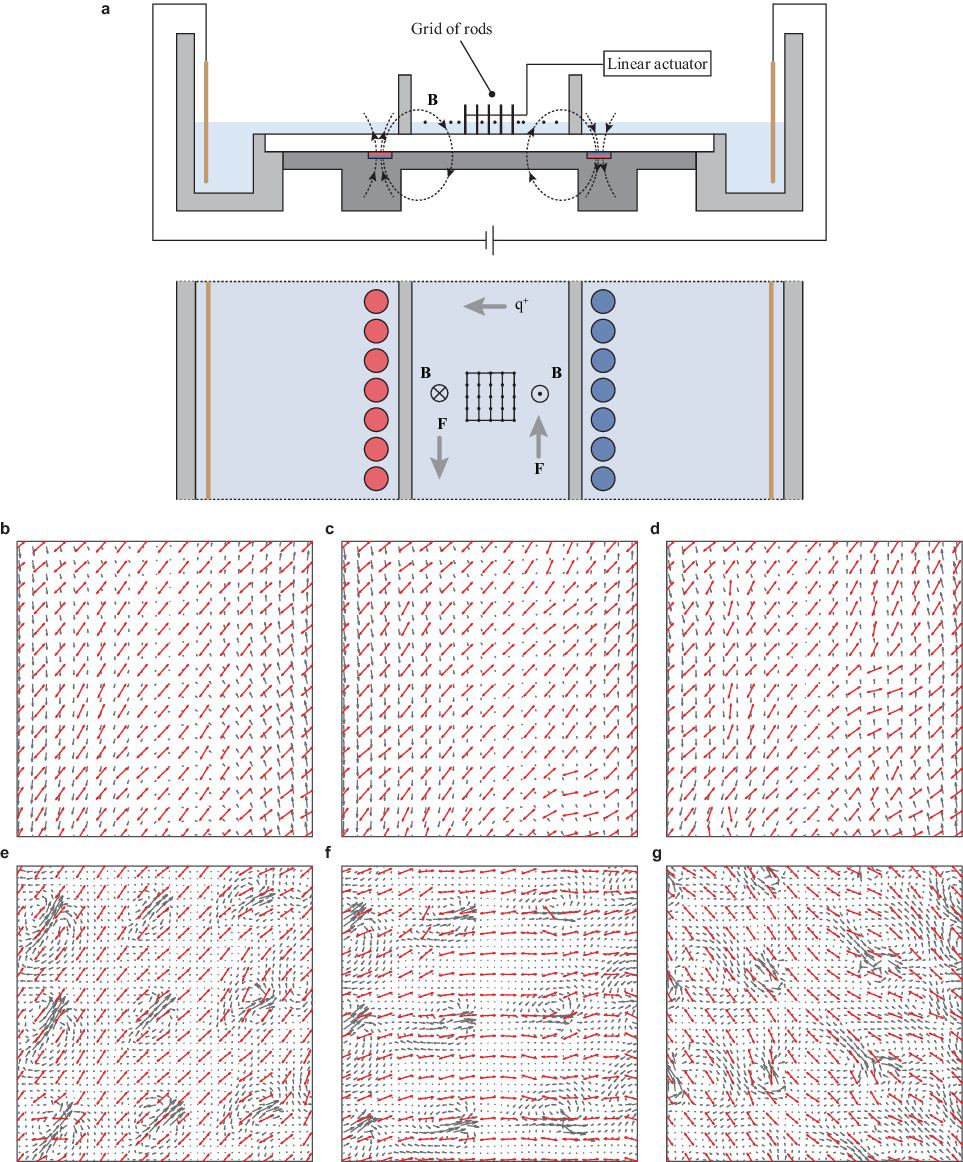}
\end{figure}
\begin{figure}
    \caption{\textbf{Perspective views of the experimental setup and the tensor geometry of the experimentally measured manipulated flow.} \textbf{a}, Perspective view of the experimental setup for testing flow manipulation via tensor geometry. \textbf{b} and \textbf{e} show the case where the array of rods were driven in a direction that formed an angle $\theta$ approximately equal to 0. \textbf{c} and \textbf{f} show the case where $\theta \approx \pi/4$. \textbf{d} and \textbf{g} show the case where $\theta \approx \pi/2$. For panels \textbf{b}, \textbf{c} and \textbf{d}, the gray arrows show the large-scale flow field after filtering as $u_i^{({\rm L})}$ and the red arrows show the directions of local extensional eigenvectors for the large-scale rate of strain tensor $s_{ij}^{({\rm L})}$. For panels \textbf{e}, \textbf{f} and \textbf{g}, the gray arrows show the small-scale residual flow field $u_i - u_i^{({\rm L})}$ and the red arrows show the directions of local extensional eigenvectors for the subgrid stress $\tau_S^{(L)}$. The filter length L for all cases were taken as L/W = 0.8. For both the velocity fields and the eigenvector fields, the arrows were down-sampled for better visualization.
    }
    \label{Ext.fig2}
\end{figure}

\begin{figure}[p]
    \centering
    \includegraphics[width=\textwidth]{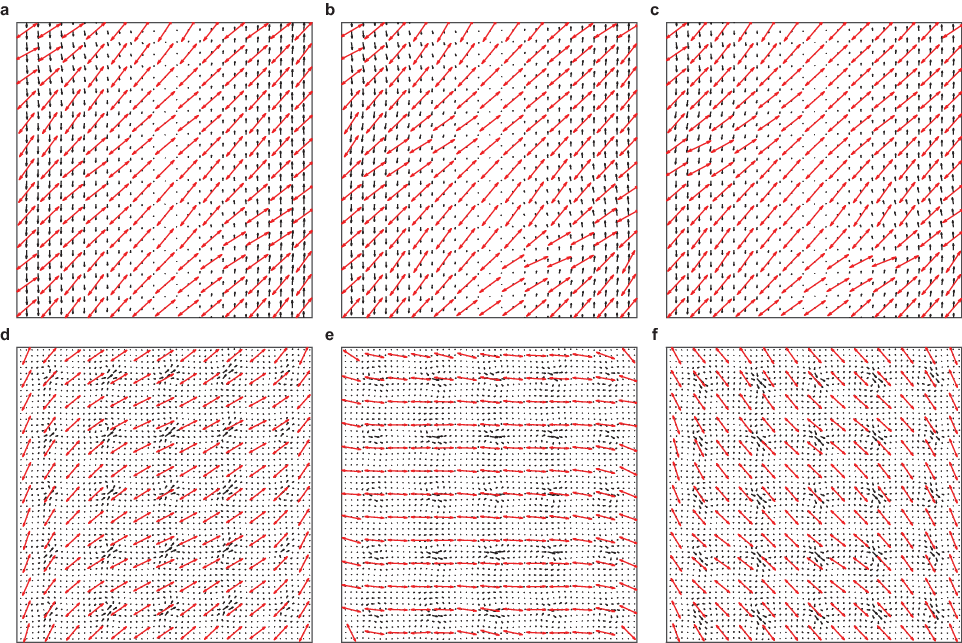}
    \caption{\textbf{Tensor geometry of the simulated manipulated flow.} \textbf{a} and \textbf{d} show the case where the array of rods was driven in a direction that formed an angle $\theta = \pi/16$. \textbf{b} and \textbf{e} show the case where $\theta =  5\pi/16$. \textbf{c} and \textbf{f} show the case where $\theta = \pi/2$. For panels \textbf{a}, \textbf{b} and \textbf{c}, the gray arrows show the large-scale flow fields after filtering ($u_i^{({\rm L})}$) and the red arrows show the directions of local extensional eigenvectors for the large-scale rate of strain tensor $s_{ij}^{({\rm L})}$. For panels \textbf{d}, \textbf{e} and \textbf{f}, the gray arrows show the small-scale residual flow field ($u_i - u_i^{({\rm L})}$) and the red arrows show the directions of local extensional eigenvectors for the subgrid stress ($\tau_S^{(L)}$). The filter length L for all cases were taken as L/W = 0.8. For both the velocity field and the eigenvector field, the arrows were down-sampled for better visualization.
    }
    \label{Ext.fig3}
\end{figure}

\begin{figure}[p]
    \centering
    \includegraphics[width=\textwidth]{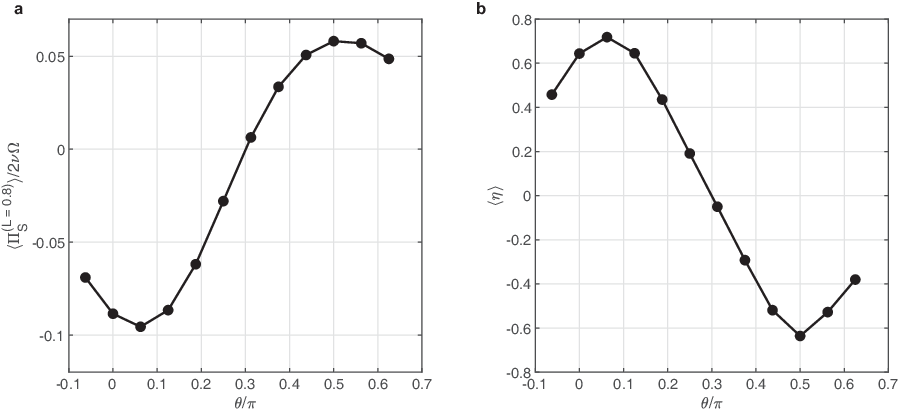}
    \caption{\textbf{Tensor geometry statistics of the manipulated flow.} \textbf{a}, A plot of the inset of Fig. 3j that shows the $\Pi^{(\rm L)}$ at L/W = 0.8 for a range of mechanical angle ($\theta$). \textbf{b}, A plot of the mean efficiency for a range of mechanical angle ($\theta$). Efficiency is defined as $\eta = \textrm{cos}(2\theta^{(\rm L)})$.}
    \label{Ext.fig4}
\end{figure}



\end{document}